\documentstyle[aps,prl,floats,graphicx]{revtex}
\begin{document}

\def\topfraction{1} \def\bottomfraction{1} \def\textfraction{0}

\draft
\title{Reentrant violation of special relativity in the low-energy
corner.}
\author{G.E. Volovik\\
Low Temperature Laboratory, Helsinki
  University of Technology\\
 Box 2200, FIN-02015 HUT, Finland\\
and\\
L.D. Landau Institute for Theoretical Physics, 117334 Moscow,
  Russia}
\maketitle
\begin{abstract}

In the effective relativistic quantum field theories the energy region,
where the special relativity holds, can be sandwiched from both the high
and low
energies sides by domains where the special relativity is violated.  An example
is provided by $^3$He-A where the relativistic quantum field theory emerges as
the effective theory. The reentrant violation of the special relativity in the
ultralow energy corner is accompanied by the redistribution of the
momentum-space
topological charges between the fermionic flavors. At this ultralow energy an
exotic massless fermion with the topological charge $N_3=2$ arises, whose
energy
spectrum mixes the classical and relativistic behavior. This effect can lead to
neutrino oscillations if neutrino flavors are still massless at this energy
scale.

\end{abstract}

{\it Introduction.} The condensed matter analogy supports an idea that the
special and general relativity might be emergent properties of quantum vacuum,
which arise gradually in the low energy corner
\cite{FrogNielBook,Chadha,PhysRepRev}. If this is true one can expect that the
Lorentz invariance of our low-energy world is violated at high energy. The
condensed matter provides examples of how this violation can occur. Here we
demonstrate one generic example which is realized in superfluid $^3$He-A, where
the effective special relativity and gravity do arise in the low energy corner
together with the chiral fermions and effective gauge fields \cite{PhysRepRev}.
It suggests that if the effective Lorentz invariance is violated in the extreme
limit of ``Planckian'' scale, it becomes violated also in the opposite extreme
limit of ultralow energy. The energy scale where the reentrant violation of the
special relativity occurs is also dictated by the ``trans-Planckian'' physics.
If there are still fermions which remain chiral and massless when approaching
this ultralow energy scale, the reentrant violation of the Lorentz invariance
leads to the crucial reconstruction of their energy spectrum. Thus the
trans-Planckian physics can be probed in the limit of low energies.

\begin{figure}[!!!t]
\centerline{\includegraphics[width=\linewidth]{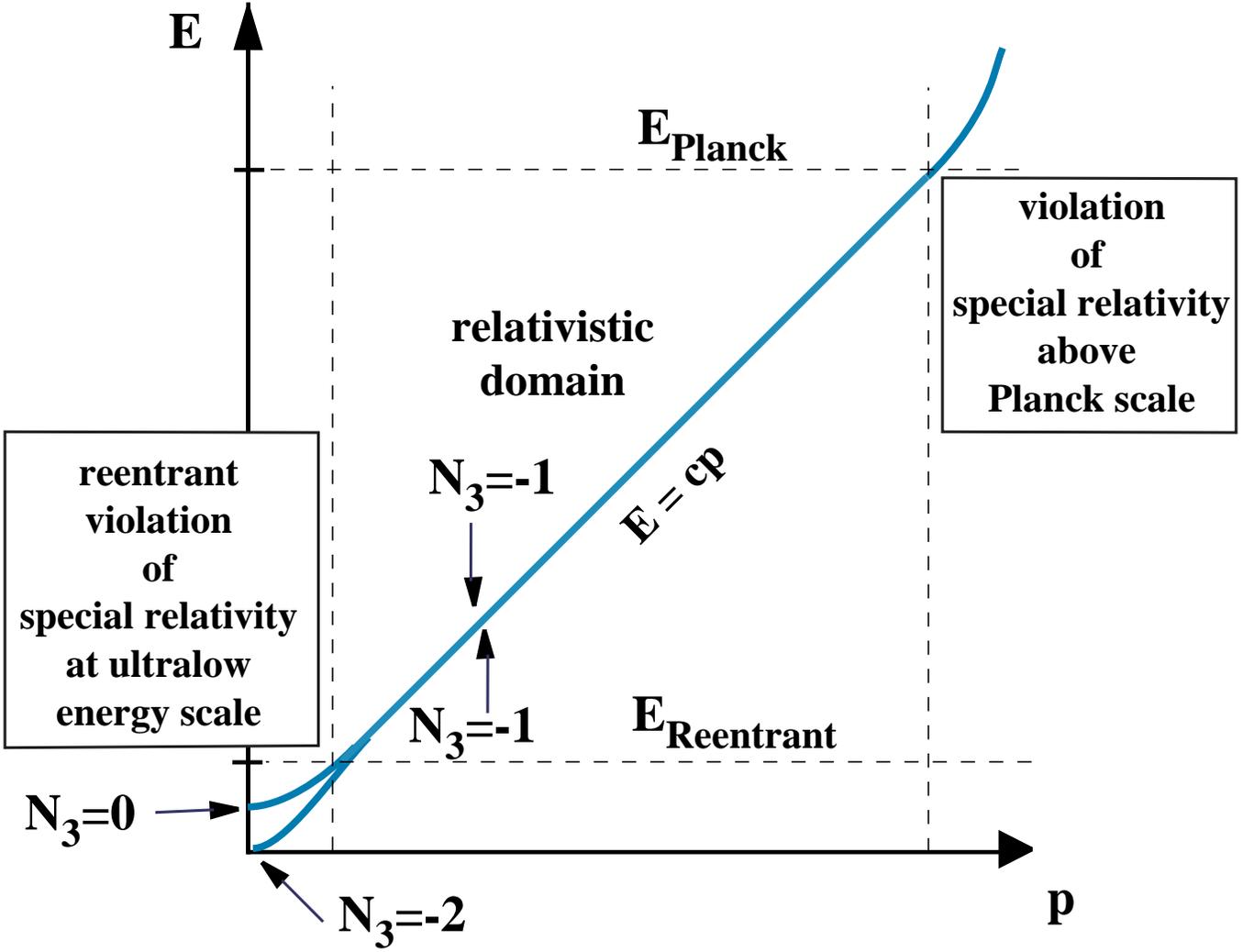}}
\bigskip
\caption[Reentrant]
    {Low energy memory of the high energy nonsymmetric physics}
\label{Reentrant}
\end{figure}

In this example two flavors of chiral left-handed fermions, the two are
hybridized producing one massive fermion with the relativistic spectrum
$E^2=c^2p^2 + m^2c^4$ and one exotic gapless fermion whose spectrum mixes the
classical and relativistic behavior:
$E^2=c^2p_\parallel^2 + (p_\perp^2/2m)^2$ (Fig.\ref{Reentrant}). Such energy
spectrum is the consequence of the nontrivial momentum space topology. The
hybridization of fermions due to violation of special relativity can
provide the
scenario for neutrino oscillations, similar to that discussed in Ref.
\cite{Glashow} where also the violation of special relativity was
considered, but
in terms of different maximum attainable velocity $c$ for different species of
neutrino.

{\it Momentum space topology and discrete symmetry between the fermions.}
The special relativity (and also the general relativity with the effective
gravitational field being one of the collective modes of the fermionic vacuum)
naturally arises in such Fermi superfluids whose fermionic quasiparticle
spectrum
contains topologically nontrivial point nodes in momentum space. Examples
are the
following superfluid phases of $^3$He:  $^3$He-A and the planar state.  In the
low energy limit, i.e. in the vicinity of a given topologically stable
point node
(the Fermi point), fermionic quasiparticles behave as chiral fermions with the
masless spectrum obeying equation
\begin{equation}
g_{(a)}^{\mu\nu}(p_\mu - p_{\mu(a)}) (p_\nu - p_{\nu(a)})=0~.
\label{MasslessSpectrum}
\end{equation}
Here $p_{\mu(a)}$ (the position of the node in the spectrum of the
$a$-th quasiparticle) and $g_{(a)}^{\mu\nu}$ are dynamic variables, which
describing the collective bosonic degrees of freedom of the vacuum. They
play  the
role of the gauge and gravity fields respectively.

In each of the two phases of superfluid $^3$He ($^3$He-A and the planar state)
there is a symmetry, which connects all the low-energy fermionic species. As a
result, the effective metric $g_{\mu\nu}$ is the same for all fermions (at
least
in equilibrium) which means that all of them have the same ``speed of light''
(i.e. the same maximum attainable speed). Moreover, for the ``perfect''
fermionic
system (for which the Lagrangian for the collective bosonic modes is
obtained by
the integration over the fermions in the vicinity of the Fermi points, see
\cite{PhysRepRev}) the bosonic fields are governed by the same effective metric
$g_{\mu\nu}$ as fermions and thus have the same speed of light. Thus the
nontrivial momentum-space topology and the symmetry between the fermions
are two
ingredients for establishing the special relativity in the low energy corner of
the effective theory.

The massless (gapless) character of the fermionic spectrum in the system with
Fermi points is protected by the nonzero value of the topological invariant
of the
ground state, which is expressed as the integral over the Green's function in
the 4D momentum-frequency space:
\begin{equation}
N_3 =
{1\over{24\pi^2}}e_{\mu\nu\lambda\gamma}~
 ~\int_{\sigma}~  dS^{\gamma}
~ {\cal G}\partial_{p_\mu} {\cal G}^{-1}
{\cal G}\partial_{p_\nu} {\cal G}^{-1} {\cal G}\partial_{p_\lambda}  {\cal
G}^{-1}~.
\label{TopInvariantMatrix}
\end{equation}
The integral here is over the surface
$\sigma$ embracing the Fermi point $p_{\mu(a)}=({\bf p}_a, 0)$
in the 4D momentum space $p_\mu=({\bf p}, p_0)$ and $p_0$ is the energy
(frequency) along the imaginary axis; ${\bf tr}$ is the trace over the
fermionic
indices.
If the value of the topological charge of the Fermi point is $N_3=+1$ or
$N_3=-1$,
then in the vicinity of this point there is a massless fermion, whose Green's
function after the proper rescaling and shifting the position of the Fermi
point
has the following form
\begin{equation}
{\cal G} =(ip_0-{\cal H})^{-1}~~,~~{\cal H}= N_3 c{\bf \sigma}\cdot {\bf p}~.
\label{GreenFunction}
\end{equation}
This is the Green's function of the left-handed (if $N_3=-1$) or
right-handed (if
$N_3=+1$) chiral fermion. Thus the topologically nontrivial
momentum-space topology automatically produces the ``relativistic''
chiral fermions as the low-energy quasiparticles, if $N_3=\pm 1$.

In the gapless superfluid phases of $^3$He the topological invariants of the
two Fermi points are different from $\pm 1$: they are $N_3=\pm 2$ in $^3$He-A
and $N_3=0$ in the planar state. In each phase, however, there is a discrete
symmetry between the fermionic species, which leads to the equal (in
$^3$He-A) or opposite (in the planar state) distribution of the
topological charges  between fermions:
$N_3=+2\rightarrow +1 +1$ and
$N_3=-2\rightarrow-1 -1$ in $^3$He-A and $N_3=0\rightarrow +1 -1$ in the planar
state. As a result each gapless fermion has unitary charge,
$N_3=-1$ or
$N_3=+1$, and thus becomes relativistic in the low-energy corner.
This again shows the importance of the symmetry between the fermions for the
special relativity to hold in the low energy corner. On the discrete symmetry
which provides the unitary charges $N_3=-1$ or $N_3=+1$ for the chiral fermions
in the Standard Model see Ref. \cite{PhysRepRev}.

{\it Condensed mater scenario of the reentrant violation of special
relativity.}
In $^3$He-A the symmetry $SO(3)_S\times
SO(3)_L\times U(1)_N$ of the normal $^3$He is broken to the little group
$U(1)_{L_z -N/2}\times U(1)_{S_z}$ whose generators are $L_z-N/2$ and $S_z$.
Here $SO(3)_S$, $SO(3)_L$ and $U(1)_N$ are correspondingly spin rotation group,
group of orthogonal coordinate transformations and the global $U(1)$ group
responsible for the conservation of the global charge -- the number of $^3$He
atoms. The corresponding $3\times 3$ order parameter matrix $A_{\mu \,i}$,
which
transforms as a vector under spin rotations $SO(3)_S$ (the first index) and
as a
vector under orbital rotations $SO(3)_L$ (the second index), is
\begin{equation}
A_{\mu \,i}=\Delta_0\hat z_\mu \left(\hat x_{i} + i \hat y_{i}\right)~.
\label{APhase}
\end{equation}
Fermionic quasiparticles living in the vacuum with such order parameter
have two
point nodes in the spectrum. In the vicinity of the Fermi point
at ${\bf p}=(0,0, p_F)$, which has with the topological charge $N_3=-2$, these
quasiparticles correspond to two chiral lefthanded ``relativistic''
fermions described by the following Bogliubov-Nambu Hamiltonian
\begin{equation}
{\cal H}_{\rm A-phase}= c_\parallel (p_z-p_F)\check \tau
_3 + {c_\perp} \sigma_z(\check \tau
_1 p_x -\check \tau
_2 p_y ) ~.
\label{BogoliubovNambuHamAPhase}
\end{equation}
Here the ``speeds of light'' propagating along and transverse to the axis $z$
are correspondingly
$c_\parallel=v_F$ and
$c_\perp=\Delta_0/p_F$; $v_F$ and $p_F$ are the Fermi velocity and
Fermi momentum in the normal $^3$He; $\Delta_0$ is the amplitude of the gap;
$\sigma_z$ is the Pauli matrix for the nuclear spin of $^3$He atom; $\check
\tau_i$ are the Pauli matrices for the Bogoliubov-Nambu spin in the
particle-hole
space.

Two projections of atomic spin $(1/2)\sigma_z=\pm 1/2$ can be considered
as two fermionic flavors. Each of two fermions has $N_3=-1$, that is why it is
chiral (left-handed) and massless, with the energy spectrum
\begin{equation}
E^2= c_\parallel^2 \tilde p_z^2+c_\perp^2 p_\perp^2~,~~N_3=-1~,
\label{APhaseEnergy}
\end{equation}
where $\tilde p_z= p_z-p_F$.
The symmetry, which couples the two flavors and forces them to have identical
topological charge $N_3=-1$ and identical ``relativistic'' spectrum, is the
discrete symmetry
$U_2$ of the order parameter in Eq.(\ref{APhase}). $U_2$ is the combined
symmetry: it the element of the
$SO(3)_S$ group, the
$\pi$ rotation of spins about, say, axis
$x$, which is supplemented by the gauge rotation $e^{i\pi}$ from the $U(1)_N$
group: $U_2=C_\pi^x e^{i\pi}$.  It is the same discrete symmetry which gives
rise to the Alice string (the half-quantum vortex, for review see
\cite{ReviewAlice}). Such symmetry came from the fundamental level well
above the
first Planck scale $E_{\rm Planck}=\Delta_0^2/v_Fp_F$ \cite{PhysRepRev} at
which
the spectrum becomes  nonlinear and thus the Lorentz invariance is violated.

However, this discrete $U_2$ symmetry is not exact in $^3$He even on the
fundamental level. This is because of the tiny spin-orbit interaction, which
slightly violates the symmetry $SO(3)_S$ under the separate rotations of spin
space. Since the $U_2$ symmetry was instrumental for the establishing of the
special relativity in the low energy corner, its violation must lead to
violation
of the Lorentz invariance and also to mixing of the two fermionic flavors
at the
very low energy determined by the tiny spin-orbit coupling.

Let us consider how all this happens in $^3$He-A.  Due to the spin-orbit
coupling
the $U(1)_{L_z -N/2}\times U(1)_{S_z}$ symmetry of $^3$He-A is not exact. The
exact symmetry now is the combined symmetry constructed from the sum of
two generators: $U(1)_{J_z -N/2}$, where $J_z=S_z+L_z$ is the generator of the
simultaneous rotations of spins and orbital degrees of freedom.
As a result the order parameter in Eq.(\ref{APhase}) acquires a small
correction
consistent with the $U(1)_{J_z -N/2}$ symmetry:
\begin{equation}
A_{\mu \,i}=\Delta_0\hat z_\mu \left(\hat x_{i} + i \hat y_{i}\right) +
\alpha\Delta_0~\left(\hat x_\mu +i\hat y_{\mu}\right)   \hat z_{i}~.
\label{APhaseCorrected}
\end{equation}
The first term corresponds to Cooper pair state with $L_z=1/2$ per atom and
$S_z=0$, while the second one is a small admixture of state with
$S_z=1/2$  and $L_z=0$. Both components have $J_z=1/2$ and thus must be present
in the order parameter. Due to the second term the order parameter is not
symmetric under the $U_2$ operation. The small parameter $\alpha\sim
\xi^2/\xi_D^2\sim  10^{-5}$ is the relative strength of the spin-orbit
coupling,
where
$\xi \sim 10^{-6}-10^{-5}$ cm is the superfluid coherence length and $\xi_D
\sim
10^{-3}$ cm is the so called dipole length characterizing the spin-orbit
coupling.

The Bogoliubov-Nambu Hamiltonian for fermionic quasiparticles in such
vacuum is now modified as compared with that in the
pure vacuum state with $L_z=1/2$   and $S_z=0$ in
Eq.(\ref{BogoliubovNambuHamAPhase}):
\begin{equation}
{\cal H}_{\rm A-phase}= c_\parallel \tilde p_z\check \tau
_3 + {c_\perp} \sigma_z(\check \tau
_1 p_x -\check \tau
_2 p_y ) + \alpha c_\perp p_z (\sigma_x \check \tau
_1  -\sigma_y\check \tau_2 )~ .
\label{BogoliubovNambuHamAPhaseCorrected}
\end{equation}
Diagonalization of this Hamiltonian shows that the small correction due to
spin-orbit coupling gives rise to the following splitting of the energy
spectrum
\begin{equation}
E_\pm^2= c_\parallel^2 \tilde p_z^2  + c_\perp^2 \left(\alpha
|p_z| \pm \sqrt{\alpha^2 p_z^2+p_\perp^2}\right)^2~ .
\label{BogoliubovNambuHamAPhaseCorrected}
\end{equation}
In $^3$He-A $p_z$ is close to $p_F$, so one can put $\alpha
|p_z|=\alpha p_F$. Then the + and - branches give the
gapped and gapless spectra correspondingly. For $p_\perp \ll
\alpha p_F$ one has
\begin{eqnarray}
E^2_{+}\approx  c_\parallel^2 \tilde p_z^2  + \tilde c_\perp^2 p_\perp^2
+\tilde m^2
\tilde c_\perp^4 ~,
\label{E+}
\\
 E^2_{-}\approx  c_\parallel^2 \tilde p_z^2  +
{p_\perp^4\over 4\tilde m^2}~,
\label{E-}\\
~\tilde c_\perp =\sqrt{2} c_\perp~,~\tilde m= \alpha ~
{p_F\over c_\perp}~,~p_\perp\ll \tilde m c_\perp ~.
\label{m}
\end{eqnarray}
In this ultra-low energy corner the gapped branch of the spectrum is
relativistic, though with different speed of light than in the intermediate
regime of Eq.(\ref{APhaseEnergy}). The gapless branch is relativistic in one
direction $E=c_\parallel |\tilde p_z|$, and is classical, $E=p_\perp^2/2\tilde
m$, for the motion in the transverse direction.

{\it Momentum space topology of exotic fermion.}
What is important here is that such splitting of the spectrum
is generic and thus can occur in other effective field theories such as
Standard
Model. This is because of the topological properties of the spectrum: the
mixing
of the two fermionic flavors occurs with the redistribution of the topological
charge between the two fermions. In the relativistic domain each of two
fermions
has the topological charge $N_3=-1$. It is easy to check that in the ultra low
energy corner it is not the case. While the total  topological charge of
the Fermi
point
$N_3=-2$, must be conserved, it is now redistributed between the fermions
in the
following manner: the massive fermion (with the energy $E_+$) acquires the
trivial
topological charge
$N_3=0$ (that is why it becomes massive), while another one (with the energy
$E_-$)  has double topological charge $N_3=-2$ (see Fig.\ref{Reentrant}). It is
important that it is now only one species of the massless fermions, which has
$N_3=-2$. Thus it cannot split into two fermions with $N_3=-1$ each.  This
exotic fermion with $N_3=-2$ is gapless because of the nonzero value of the
topological charge, but the energy spectrum of such fermion is not linear. That
is why it cannot be described in the relativistic language.

In the same way as
the $N_3=\pm 1$ fermions are necessarily relativistic and chiral in the
low-energy corner, the fermions with higher $|N_3|$ are necessarily
non-relativistic. The momentum space topology which induces the special
relativity if $|N_3|=1$, becomes incompatible with the relativistic invariance
if $|N_3|>1$, and the latter is violated.  Properties of the fermionic systems
with multiple zeroes, $|N_3|>1$, including the axial anomaly in its
non-relativistic version were discussed in Ref.\cite{VolovikKonyshev}.

The energy scale at which the splitting of the energy spectrum occurs is
$E_{\rm Reentrant}=\alpha
\Delta_0$ which is much less than the first Planck level in $^3$He-A, $E_{\rm
Planck}=\Delta_0^2/v_Fp_F$. Thus the relativistic region for the $^3$He-A
fermions, $E_{\rm Reentrant} \ll E\ll E_{\rm Planck}$, is sandwiched from
both the
high and low energies by the nonrelativistic regions.

{\it Discussion.} The above example of $^3$He-A shows that the discrete
symmetry
between the fermions together with the momentum-space topology garantee that
massless fermions obey the special relativity in the low-energy corner. If the
discrete symmetry is approximate, then in the ultralow energy corner the
redistribution of the momentum-space topological charges occurs between the
fermions with appearance of the higher topological charge $|N_3|>1$. This
topological transition leads to  the strong modification of the energy spectrum
which becomes essentially non-relativistic.

In principle, such topological transition with appearance of
the exotic fermions with $N_3=\pm 2$ can occur in the relativistic theories
too,
if these theories are effective. In the effective theory the Lorentz invariance
(and thus the special relativity) appears in the low-energy corner as an
emergent
phenomenon, while it can be violated at high energy approaching the Planck
scale. At low energy the fermions are chiral and relativistic, if there is a
symmetry between the flavors. If such symmetry is violated,
spontaneously or due to the fundamental physics above the Planck scale, then in
the extreme low energy limit, when the asymmetry between the fermionic flavors
becomes important,  the system remembers its high-energy nonrelativistic
origin.
The rearrangement of the topological charges $N_3$ between the fermionic
species
occurs and the special relativity disappears again.

This scenario can be applied to the massless neutrinos. The violation of the
horizontal symmetry between the left-handed neutrino flavors can lead to the
violation of Lorentz invariance at the very low energy. If the neutrinos
remain
massless at this ultralow energy scale, then below this scale two flavors,
say, electronic and muonic left-handed neutrinos each with
$N_3=-1$, hybridize and produce the $N_3=0$ fermion with the gap and the exotic
gapless $N_3=-2$ fermion with the essentially nonlinear non-relativistic
spectrum.
This is another example of violation of the special relativity, which can also
give rise to the neutrino oscillations. The previously considered effect of the
violation of the special relativity on neutrino oscillationis was related
to the
different speeds of light for different neutrino flavors \cite{Glashow}
(the related effect is the violation of the weak equivalence principle:
different
flavors are differently coupled to gravity
\cite{VEPNeutrino}).

This work  was supported in part by the Russian Foundation for Fundamental
Research and by European Science Foundation.

\end{document}